# Picosecond lifetimes of hydrogen bonds in the halide perovskite CH$_3$NH$_3$PbBr$_3$

Alejandro Garrote-Márquez,[a] Lucas Lodeiro,[b] Norge Cruz Hernández,[a] Xia Liang,[c] Aron Walsh,[c] and Eduardo Menéndez-Proupin[a*]

[a] *Departamento de Física Aplicada I, Escuela Politécnica Superior, Universidad de Sevilla, Seville E-41011, Spain*

[b] *Departamento de Química, Facultad de Ciencias, Universidad de Chile, Las Palmeras 3425, Ñuñoa 7800003, Santiago, Chile*

[c] *Thomas Young Centre and Department of Materials, Imperial College London, London SW7 2AZ, UK*

*Corresponding author. E-mail: emenendez@us.es

**Abstract**

The structures and properties of organic-inorganic perovskites are influenced by the hydrogen bonding between the organic cations and the inorganic octahedral networks. This study explores the dynamics of hydrogen bonds in CH$_3$NH$_3$PbBr$_3$ across a temperature range from 70 K to 350 K, using molecular dynamics simulations with machine-learning force fields. The results indicate that the lifetime of hydrogen bonds decreases with increasing temperature from 7.6 ps (70 K) to 0.16 ps (350 K), exhibiting Arrhenius-type behaviour. The geometric conditions for hydrogen bonding, which include bond lengths and angles, maintain consistency across the full temperature range. The relevance of hydrogen bonds for the vibrational states of the material is also evidenced through a detailed analysis of the vibrational power spectra, demonstrating their significant effect on the physical properties for this class of perovskites.

## 1. Introduction

Tin-lead halide perovskites have become important for use in optoelectronic devices such as solar cells and LEDs, showing very high energy conversion efficiency.[1–6] In the first publication on this field in 2009, Kojima *et al.*[7] reported MAPbBr$_3$ and MAPbI$_3$ (MA=CH$_3$NH$_3$) based solar cells that showed high photovoltage of 0.96 V and 0.61 V with external quantum conversion efficiencies of 65% and 45%, respectively.

In the subsequent years, researchers have made significant progress in improving the efficiency and performance of tin-lead halide perovskite solar cells.[8–11] There are significant advances in the development of new synthesis and manufacturing methods, which allow to achieve greater stability and reproducibility.[12,13] Research has also been conducted to understand and control the mechanisms of photogeneration. In terms of applications, solar cells based on tin-lead halide perovskites have demonstrated energy photo-conversion efficiencies (PCE) over 26%, approaching the performance of crystalline silicon solar cells.[9,14,15] In addition, efficient LEDs based on tin-lead halide perovskites, which can emit light in different colours, have been developed.[16–18] However, despite the progress achieved so far, there are still technical and scientific challenges for industrial use, one of the main ones being the long-term stability of these materials.[19–22]

The structure of MAPbBr$_3$ offers a combination of chemical and physical characteristics that can be exploited to enhance its long-term stability.[23–25] Due to the composition and size of the ions (Pb$^{2+}$) and bromine (Br$^-$), it is possible to form a stable crystalline lattice. The relatively large size of the methylammonium ion in the structure helps to maintain the stable crystalline lattice and reduces the probability of structural degradation.[26] Furthermore, lead halide perovskites like MAPbBr$_3$ tend to exhibit good thermal and environmental stability compared to other perovskites, as well as lower sensitivity to water and oxygen.[12,23]

In the structure of the perovskite MAPbBr$_3$, the lead ions (Pb$^{2+}$) occupy the sites of a simple cubic lattice, and the bromine ions (Br$^-$) occupy the positions between neighbouring lead ions. Thus, each lead is bound to six bromines forming an octahedron with lead at the centre and bromines at the corners, which connect adjacent octahedra. The methylammonium cations (CH$_3$NH$_3^+$ or MA$^+$) are found in the cavities formed between corner-sharing PbBr$_6$ octahedra. Each MA$^+$ cation is surrounded by 12 bromine and 8 lead ions, forming a cube around the methylammonium cation, as seen in Figure 1.

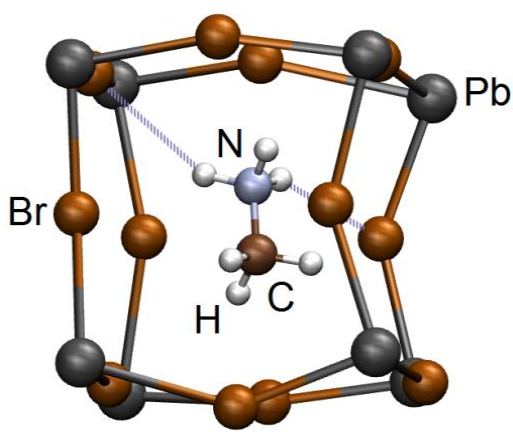

**Figure 1.** Representation of the MAPbBr$_3$ perovskite structure. Hydrogen bonds are indicated by thin dashed lines. Image created with VESTA.[27]

Temperature changes cause solid-state phase transitions in MAPbBr$_3$.[28–30] At low temperatures (below 145 K), MAPbBr$_3$ adopts an orthorhombic crystal structure. For temperatures in the range of 145 – 235 K, MAPbBr$_3$ adopts a tetragonal structure, and for higher temperatures the crystal symmetry becomes cubic. The latter is important for technological applications at room temperature. On the other hand, the stability and durability of hydrogen bonds (HBs) in lead halide perovskites are important since they play a role in the structure stability and dynamics, and hence the electrical, optical and mechanical properties of these materials. Specifically, there are three notable aspects:

1. The stability of the HBs determines the general stability of tin-lead halide perovskites. If the HBs are weak and prone to breakage, the perovskite structure can become unstable and break down easily.[31]
2. The HBs modify the distortion of $PbBr_6$ octahedra, which in turn determine the absorption and emission bands in the electromagnetic spectrum.
3. The electrical conductivity could be modified due to the mobility of ions under the influence of HBs,[32–34] which can allow the development of materials with ionic transport properties.

The existence of HBs is based on certain characteristics of hydrogen and the atoms it can form bonds with. These characteristics are defined by the International Union of Pure and Applied Chemistry (IUPAC),[35] and are as follows:

1. <u>Electronegativity difference</u>: Hydrogen must be covalently bonded to a highly electronegative atom (X), such as oxygen (O), nitrogen (N) or fluorine (F). These atoms can attract the electrons of the bond towards them, generating a negative and positive partial charge on the electronegative and hydrogen atoms, respectively.
2. <u>Bond length</u>: The length of the donor-hydrogen covalent bond (X-H) usually gets larger upon formation of HB. This is due to the electrostatic interaction between the positive partial charge of the hydrogen and the negative partial charge of the acceptor atom (Y).
3. <u>Bond angle</u>: The HB angle is close to 180º, i.e. the atoms involved (X, H, Y) lie close to a straight line.
4. <u>Forces of attraction</u>: HBs are stronger than Van der Waals forces, but weaker than covalent or ionic bonds. This relatively strong electrostatic attraction results in characteristic properties of molecules containing them, such as higher boiling and melting points than expected for compounds of similar mass.

In $MAPbBr_3$, two types of X─H···Br HBs are possible, with X=C or N. The role of carbon as the X donor is arguable due to the smaller electronegativity of carbon compared with oxygen, nitrogen, and fluorine. However, carbon fulfils the current IUPAC criteria, and its existence is also supported by non-covalent interaction analysis.[36–38] Furthermore, Figures S1 and S2 in the Supporting Information show a small enlargement of the X─H bond length correlated with decrease of the H···Br distance.

In a recent ab initio molecular dynamics (AIMD) study[37] the geometric conditions for the existence of HBs at the molecular level were defined. However, these were available only for a temperature of 350 K. In this work, we present new molecular dynamics (MD) simulations performed with a machine learning force field (MLFF).[39] The animation videos V1 to V6 in the Supporting

Information show the dynamics of HBs for different temperatures. The bonds tend to break due to the translation of the methylammonium (MA) cation and the change in orientation of the C—N vector. To a lesser extent, this occurs due to the rotation of the $CH_3$ or $NH_3$ groups. Generally, when a bond breaks, it does not reform but rather establishes a new bond with another Br. At low temperature (125 K), the ammonium group forms three N—H⋯Br bonds most of the time. These bonds frequently involve two Br atoms on opposite edges of the same face and another Br on an edge of another face. When a bond breaks, another bond quickly forms with a different Br, typically associated with a slight change in the orientation of the C—N vector or a rotation of the ammonium group around this vector. None of the observed MA molecules flip over during this time interval. At high temperature (325 K), the movements are broader, particularly the translation motion of the MA with large amplitude, superimposed with rotations around the C—N axis and changes in orientation. The breaking of hydrogen bonds (HBs) seems to be associated with the translational movement of the MA. The rotations are broader, but the cations remain relatively stable, and during the 2 ps of this animation, only one MA is seen to flip over. Observing the animations at intermediate temperatures, we can affirm that the bonds become more stable as the temperature decreases, particularly below 270 K, although at no temperature are the bonds totally stable. To quantify these effects, in this article, we conduct a statistical study of the lifetimes and other aspects of the HBs. Based on these molecular dynamics simulations, we have extended the previous statistical study of HBs in $MAPbBr_3$ to a wide temperature range, i.e., from 50 K to 350 K. The statistical functions related to HBs here used include combined (radial and angular) distribution functions, autocorrelation functions, lifetimes (LTs), and frequency distributions. The availability of LTs as functions of temperature allows to explore Arrhenius type behaviour and to obtain activation energies associated with HB breaking.

## 2. Methods
### 2.1 MLFF molecular dynamics

The interatomic forces were computed using the machine-learning force fields[40–42] based on atom-centred radial and angular descriptors as implemented in VASP.[43] The force field was trained following the workflow proposed by Liang X *et al.*[39] Training data is collected with separate MD calculations with on-the-fly learning mode enabled, starting from 100, 160, 210 and 350 K, respectively. In these MD calculations, an isothermal-isobaric ensemble with atmospheric pressure is first adopted, with a 2 × 2 × 2 supercell and a timestep of 0.5 fs. The equilibrated atomic positions and cell sizes are then applied to a second isothermal-isochoric ensemble using the same settings, where the DFT frames are collected. Langevin thermostat is adopted for both MD calculations. For the DFT calculations in the on-the-fly training, $r^2SCAN$ exchange-correlation functional[44] is selected, the plane-wave basis set cutoff energy is 500 eV, and the electronic convergence threshold is $10^{-5}$ eV. A general force field is then trained on the collected

DFT frames from the four temperatures. The detailed force field training procedures can be found in Ref. 39.

The MD simulations intended to obtain the HB statistics have been performed in the canonical ensemble with the Nosé thermostat,[45] using the force field trained from all the collected DFT frames in the previous section. The simulation cell contains 2592 atoms, which corresponds to a $6 \times 6 \times 6$ supercell of the cubic phase primitive cell. The dimensions of the simulation cell for each temperature were determined as the average values in a previous simulation with the isothermal-isobaric ensemble. The equations of motion were integrated with a time step of 0.5 fs, while the coordinates were saved every two-time steps. The equations of motion need to be integrated with such small time steps to obtain accurate vibrational frequencies.[46]

## 2.2 Hydrogen bonding analysis

To characterize the X─H···Br HB (X=C or N) we have used a combined distribution function (CDF) of the distance H─Y and the angle Br─H─X. This CDF describes the joint probability of finding a hydrogen-acceptor distance, simultaneously with a certain angle Br─H─X, in the ensemble generated from the MD. These CDFs have been computed using the TRAVIS code.[47] Figure 2 shows the CDF with the HB length plotted on the horizontal axis, while the angle of the HB is displayed on the vertical axis. The red spots indicated the ranges of maximal probability to find the distance H─Br and the angle Br─H─X in the ensemble generated by molecular dynamics.

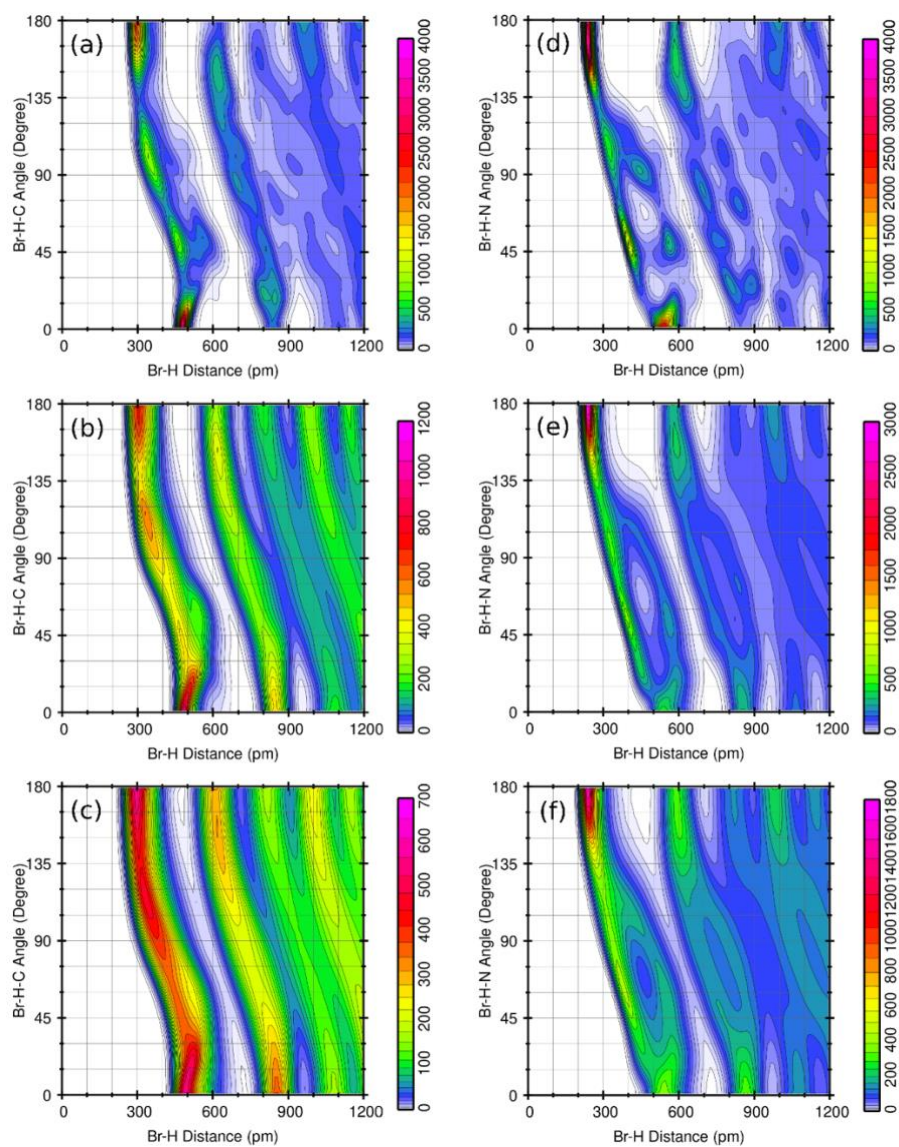

**Figure 2.** Combined distribution functions of Br—H distance in pm (horizontal axes) with ∢(Br − H − X) angle (X=C or N) in degrees (vertical axes) for temperatures at 70 K (a and d), 175 K (b and e) and 350 K (c and f).

The 2D contour plots in Figure 2, show zones of high concentration (red and purple colours), indicating the presence of HBs, when the distances are shorter than 300 pm, and the angles are close to 180°. These graphs provide information on the geometric characteristics necessary for the formation of the HB. This demonstrates a correlation between both variables, information that standard 1D histograms cannot provide. In Garrote-Márquez *et al*,[37] it was shown, based on the CDFs, that instantaneous N—H⋯Y HBs (Y=Br or I) in the MD ensemble are characterized by simultaneous fulfilment of the conditions $d(N − H) < 3$ Å, and $135º < ∢(I − H − N) < 180°$. Furthermore, C—H⋯Y HBs are revealed by modifying the distance condition to $d(C − H) < 4$ Å. The analysis of the CDFs here presented suggests that these geometric conditions remain valid for all temperatures smaller than 350 K, which the previous study was performed for.

On the other hand, to determine the LTs of HBs, it is not sufficient to consider just a single trajectory, as a molecular dynamics simulation has chaotic behaviour. The approach would be to perform a simulation where many HBs have formed and broken a significant number of times, and then obtain the time correlation function.[48]

$$C_C^{HB}(t) = \frac{1}{N_1 N_2} \sum_{i=1}^{N_1} \sum_{j=2}^{N_2} \int_0^\infty \beta_{ij}(t') \tilde{\beta}_{ij}(t'+t) dt', \qquad (1)$$

where $\beta_{ij}(t') = 1$ if at instant $t'$ there is a HB between atoms $i$ and $j$ (one halide and one hydrogen, and implicitly C or N as discussed above) or zero otherwise. The function $\tilde{\beta}_{ij}(t'+t)$ is zero if the HB breaks at any instant between $t'$ and $t'+t$, or one if the HB keeps formed all time in this interval. The LT of the HB can be obtained as:

$$\tau = 2 \int_0^\infty C_C^{HB}(t) dt \qquad (2)$$

The TRAVIS code fits the correlation function $C_C^{HB}(t)$ to a sum of exponential functions using a least squares procedure,[47] and performs the integration analytically. The experience indicates that to obtain converged values of lifetimes, the MD simulation time should be larger than $10\tau$. This requirement imposes that the dynamics for low temperatures require longer simulation times. All these calculations have been performed for temperatures ranging from 70 K to 350 K, which includes the three crystal phases of MAPbBr$_3$. The sampling time has been 40 ps for $T > 125$ K and 60 ps for $T \leq 125$ K.

To obtain the activation energies associated with HB breaking, we have fitted the inverse of lifetimes with the Arrhenius equation and with the Eyring equation.

## 3. Results

### 3.1. Verification of geometrical condition from the CDF

Figure 2 depicts the prevalence of HBs using CDFs, as functions of the Br—H distance and Br—H—X angle (X = C or N). The highest values are found in the purple-shaded regions, as indicated in the scale. It is observed that at higher temperatures, the red-purple shaded areas expand, initially suggesting that the existence of HBs is greater at elevated temperatures. However, the scale also changes as the temperature increases, decreasing the maximum count values from 4000 at 70 K to 700 at 350 K. Due to this circumstance, a direct comparison of CDFs at different temperatures cannot be made. Nevertheless, it does serve to confirm the geometric conditions in which the existence of HBs is more likely. The regions of higher count can be determined by the simultaneous conditions of $d < 3$ Å, and $135º < \sphericalangle(\text{Br} - \text{H} - \text{N}) < 180°$ when X = N, and of $d < 4$ Å, and $135º < \sphericalangle(\text{Br} - \text{H} - \text{C}) < 180°$ when X = C.

## 3.2. Time correlation functions

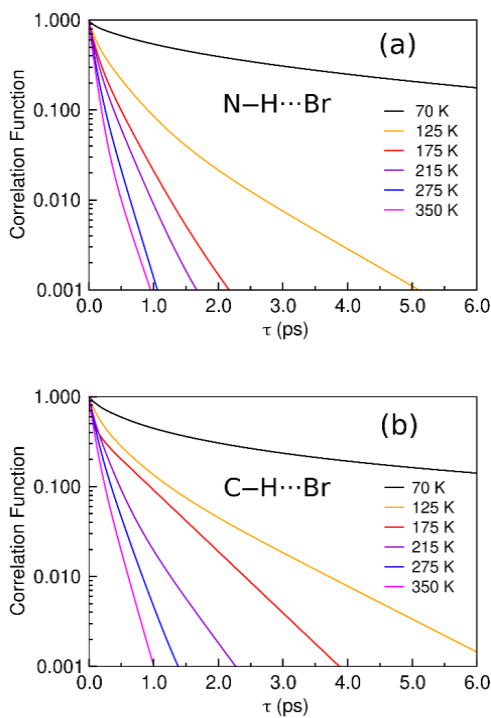

**Figure 3.** Autocorrelation functions of N─H···Br (a) and C─H···Br (b) HBs for the orthorhombic (70 K), tetragonal (125, 175, 215 K) and cubic (275 and 350 K) phases.

Once the geometric conditions that define the HBs have been confirmed to be valid for the full range of temperatures, the HB dynamics can be characterized using the time autocorrelation functions defined in equation (1), as well as the LT defined by equation (2). Figure 3 shows the correlation functions, which decay faster as the temperature increases, as expected. It can be appreciated, thanks to the logarithmic vertical scale, that none of the correlation function decay as a single exponential function. The decay is multiexponential, and in all the studied cases, they are very well fitted using either three or four exponential functions. The LT derived from the HBs, are shown in Table 1. The number of exponential functions used in each case is also shown in Table 1. For all cases the goodness of fit parameter $R>0.99997$.

**Table 1.** Lifetimes at C─H···Br and N─H···Br HBs derived from the HB continuous time correlation functions. The number of exponential functions $N_f$ is given.

| T (K) | C─H···Br | | N─H···Br | |
|---|---|---|---|---|
| | τ (ps) | $N_f$ | τ (ps) | $N_f$ |
| 70.0 | 7.6408 | 4 | 6.6992 | 4 |
| 90.0 | 3.4374 | 4 | 2.7417 | 4 |
| 100.0 | 2.5533 | 4 | 2.0081 | 4 |
| 110.0 | 1.5975 | 4 | 1.4053 | 4 |
| 125.0 | 0.7176 | 4 | 0.9616 | 4 |
| 150.0 | 0.5085 | 3 | 0.6800 | 4 |
| 175.0 | 0.3878 | 4 | 0.5150 | 3 |
| 200.0 | 0.3409 | 4 | 0.4166 | 4 |
| 215.0 | 0.3044 | 4 | 0.3830 | 3 |
| 235.0 | 0.2464 | 4 | 0.3383 | 4 |
| 250.0 | 0.2349 | 3 | 0.3188 | 4 |
| 275.0 | 0.2082 | 3 | 0.2787 | 4 |
| 300.0 | 0.1933 | 3 | 0.2532 | 4 |
| 325.0 | 0.1774 | 3 | 0.2267 | 3 |
| 350.0 | 0.1635 | 3 | 0.2035 | 4 |

It is observed that the LTs increase as the temperature decreases. Furthermore, the LTs of N─H···Br bonds are longer than the LTs of C─H···Br bonds in the cubic and tetragonal phases. However, in the orthorhombic phase the LTs the opposite trend is observed. Nevertheless, this latter comparison must be carefully evaluated, as the comparison of the LTs of C─H···Br and N─H···Br is biased due to the difference in the distance cutoffs. Hence, this comparison can only be regarded as a trend. Let of note that for the LT to be accurately computed, the TRAVIS code recommends that the MD simulation must be ten times longer than the LT. We have verified that this condition is necessary. For T=125 K, a sampling time much larger than 10τ, 60 ps, was needed to obtain a converged value of the LT. This anomaly if probably related to the tetragonal/orthorhombic phase transition. The LTs for 70 K are slightly larger than one tenth of the simulation time, but they still fit into the Arrhenius trend, as is discussed in the next section. For 50 K, LTs of 16 ps and 18 ps were obtained for N─H···Br and C─H···Br bonds, respectively, but these values seem inaccurate because they fall well below the Arrhenius trend and are also much larger than 10% of the simulation time. Extending the simulation time to obtain accurate LTs for 50 K would have required excessive storage of atomic coordinates, hence it was not done.

### 3.3. Arrhenius plots

Figure 4a depicts the relationship between the LT of N─H···Br HBs and the inverse temperature. The computed LTs, marked with purple crosses, suggest a linear dependency when plotted on a logarithmic scale of LTs against the inverse of temperature, which is characteristic of Arrhenius equation plots. The Arrhenius-type behaviour of LTs refers to the temperature dependence of the reaction rate of HB breaking. According to the Arrhenius equation,[49,50] the reaction rate $k$ can be expressed as:

$$k = Ae^{-\frac{E_a}{k_B T}},$$

where, $A$ is a constant pre-exponential factor, $E_a$ is the activation energy of the reaction, $k_B$ is the Boltzmann constant, and $T$ is the absolute temperature. In our context, the reaction is the breaking of the HB, and the reaction rate is the inverse of lifetime $k = 1/\tau$. Hence, the activation energy can be obtained from the least-square fit of the linearized equation of $\ln \tau$ vs $1/T$

$$\ln \tau = \frac{E_a}{k_B}\frac{1}{T} - \ln A.$$

The slope of the fitted line, multiplied by $k_B$, provides the activation energy for the dissociation of the HBs. The different activation energy values for three distinct phases, α (cubic) and β (tetragonal) and γ (orthorhombic), denoted by $E_\alpha$, $E_\beta$ and $E_\gamma$, respectively, reveal subtle differences in the stability of bonds under different phase conditions. Notably, the α phase exhibits a slightly higher activation energy than the β and γ phases. Nonetheless, the linear trend is clear for the three phases, and the correlation coefficients are 0.994, 0.999, 0.996 for the α, β and γ phases, respectively. There is one notable outlier from the linear trend, which has been dismissed from the linear fits. At 235 K, MAPbBr$_3$ is in α phase but close to the α–β transition. The HB lifetime at 235 K is smaller than it should be, and it seems to fit better with the line of the β phase. This anomaly may be related to the fact that structural parameters show a phase with mixed cubic-tetragonal character.[39]

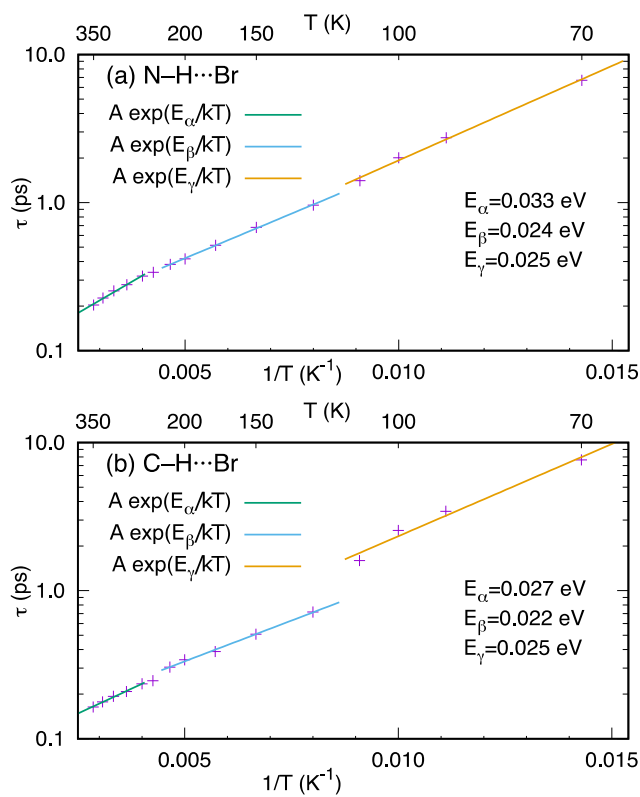

**Figure 4.** Arrhenius equation plot between the LT of N-H···Br and C-H···Br HBs and the inverse temperature for MAPbBr$_3$.

A similar analysis for C─H···Br HBs is shown in Figure 4b. Comparatively, the slopes of the lines indicate different activation energies for C─H···Br and N─H···Br HBs. Figure 4 indicates a slightly lower activation energy, for the cubic and tetragonal phases, for C─H···Br than the N─H···Br bonds, consistent with the idea that N─H···Br HBs are stronger. Moreover, the lifetime of for C─H···Br HBs shows the same kind of deviations at 235 K that was mentioned for N─H···Br bonds. Moreover, there is marked jump between temperature 125 K and 110 K, coincident with the transition between the tetragonal and orthorhombic phases.

The Eyring equation[51] fits the data worse than the Arrhenius equation. The Eyring-type plots in Supporting Information show visually important deviations from linearity, and the correlation coefficients are smaller than in the Arrhenius case.

### 3.4. Neighbour analysis

Figure 5 shows, as a function of temperature, the statistical distribution of the number of bromine ions linked to MA cations through N─H···Br or C─H···Br HBs. These numbers correspond to the number of HB per MA cation, disregarding the very rare configurations where some Br establish more than one HB with the same cation (none has been seen in the animations). At low temperatures, MA cations establish N─H···Br HBs with three Br (~80% of configurations at 50 K), or two Br (20% at 50 K), while other coordination numbers are negligible. At high temperatures, twofold coordination overcomes threefold coordination (51% vs 28% at 350 K). The crossover between threefold and twofold coordination through N─H···Br HBs takes place near 175 K. Moreover, there is a pronounced change in slope at 125 K, coincident with the orthorhombic/tetragonal phase transition. A similar, although less pronounced change, takes place near the tetragonal/cubic phase transition, at 235 K. The likelihood of MA forming just one HB is negligible at low temperature, but it begins to rise after 125 K, reaching 18% at 350 K. The statistics of C─H···Br HBs has similarities and differences. Threefold coordination is the most frequent one in the full temperature range. Twofold coordination is the second in importance. Like for N─H···Br HBs, there is a pronounced change in slope around 125 K. Surprisingly, fourfold coordination is the third most frequent case, with a 10% share. We think this abundance of high coordination is a consequence of the enlarged C─H···Br distance cutoff compared with the N─H···Br distance cutoff, and the confinement of MA in the cuboctahedra cages. The N─H···Br bonds are stronger than the C─H···Br ones, as indicated by shorter bond distances and the effect on the power spectra discussed below. This can be summarized as follows

- HBs in N─H···Br are potentially stronger but are more impacted by rising temperatures.
- HBs in C─H···Br maintain a certain stability throughout the analysed temperature range.

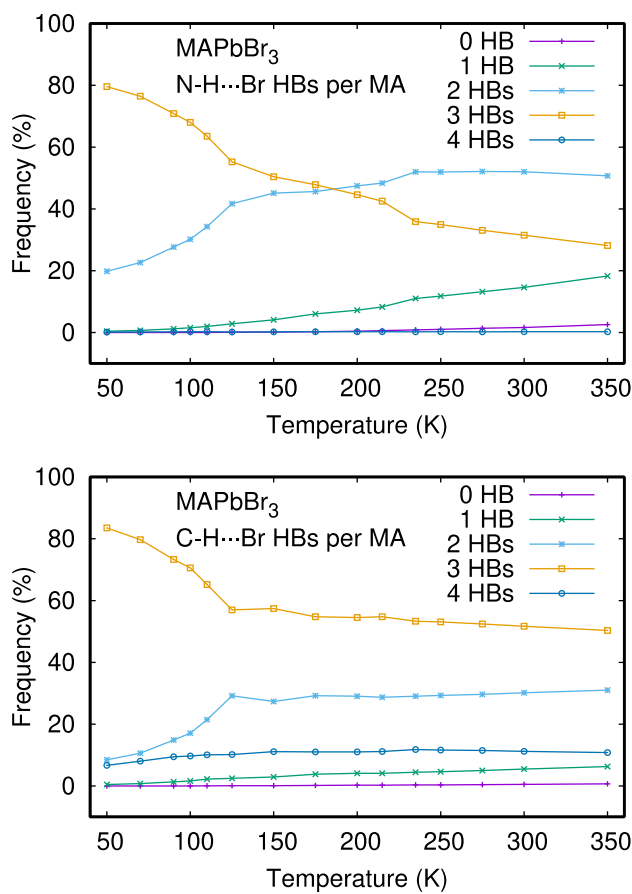

**Figure 5.** Distribution of the number of bromine ions linked to MA cations through N─H···Br or C─H···Br HBs, as function of temperature.

We could attempt to estimate the HB energies by comparison with the obtained activation energies. Under this assumption, the energy per cation can be modelled as follows.

$$E/MA = \sum_{n=1}^{3} n_i p_i E_a,$$

where $n_i$ is the number of bonds, $p_i$ is the probability (frequency/100%) and $E_a$ is the Arrhenius activation energy. As a result, we obtain that the HB energy per cation for 250 K in the N-H···Br HB is 0.073 eV, for 150 K it is 0.059 eV, and for 70 K it is 0.069 eV.

A recent calculation[31] of the HB energy per cation of MAPbBr$_3$ returned a value of 0.26 eV. This calculation is based on static relaxed structure, and the HB energy was obtained from a partition scheme of the electrostatic energy. With our molecular dynamics – Arrhenius approach, we have obtained 0.069 eV for 70 K, and 0.075 eV for an ideal structure with 100% of cations with three HBs. Our values are about one third/fourth of the former values, which is understandable due to the differences of the calculation approach, letting aside the difficulty in identifying activation energies with HB energies.

### 3.5. Vibrational power spectra

Figure 6 shows the power spectrum for selected temperatures in the range of wavenumbers that encompasses the N—H and C—H bond stretching. The power spectrum provides a vibrational density of states that includes anharmonic effects.[46] Its interpretation is aided by the knowledge of the harmonic normal modes of an isolated MA$^+$ cation, shown in Table 2. These modes have been computed ab initio with the same functional employed to generate the MLFF.[39] Modes 1 and 2 are degenerate with E symmetry, and the same is true for modes 4 and 5. The small splitting observed in Table 2 is a consequence of the breaking of rotational symmetry in the calculations with periodic boundary conditions.

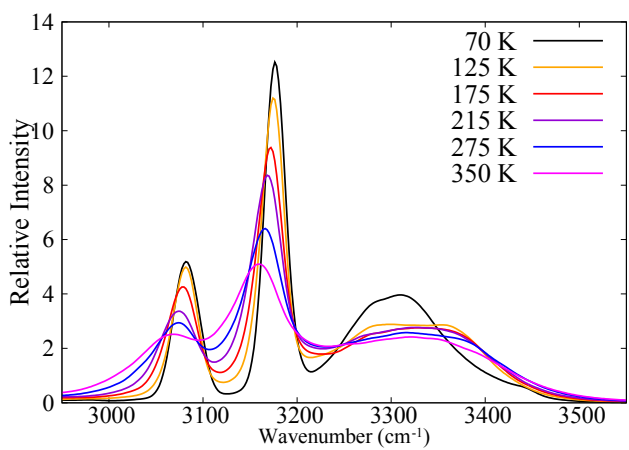

**Figure 6.** Power spectrum in the region of N—H and C—H bond stretching as a function of temperature.

Comparing the curves of Figure 6 with the wavenumbers in Table 2 we assign the two narrow peaks at the left side to the CH$_3$ symmetric and asymmetric stretching. The isolated cation normal mode frequencies are modified by the interaction with the environment and anharmonicities. The frequencies of these modes (peak maximums) in the perovskite environment vary from 3083 to 3061 cm$^{-1}$ and 3179 to 3158 cm$^{-1}$ when temperature raises from 50 to 350 K. We conclude that normal modes of C—H bond stretching are weakly affected by the environment and anharmonicity. In contrast, the higher frequency modes, associated with the N—H bond stretching, are strongly broadened and redshifted in comparison with the isolated cation modes. This is a signature of the HBs. The shape of this band is rather constant for the cubic and tetragonal phases, but it changes noticeably for temperatures below 125 K, in the orthorhombic phase.

**Table 2**. Normal modes of the N—H and C—H bond stretching in the isolated MA$^+$ cation.

| Mode | Wavenumber (cm$^{-1}$) | Symmetry | Description |
|---|---|---|---|
| 1 | 3491 | E | NH$_3$ asym. stretching |
| 2 | 3489 | | |
| 3 | 3413 | A$_1$ | NH$_3$ sym. stretching |
| 4 | 3199 | E | CH$_3$ asym. stretching |
| 5 | 3198 | | |
| 6 | 3090 | A$_1$ | CH$_3$ sym. stretching |

## 4. Discussion and conclusions

In this study it has been demonstrated that the lifetimes of the HBs in MAPbBr$_3$ are in the picosecond regime, showing that greater thermal energy facilitates overcoming the energy barrier for the dissociation of the HBs. This is manifested in a decrease in the lifetimes of the HBs with an increase in temperature, following Arrhenius behaviour. The specific values of the activation energy, $E_a$, for different phases and types of HBs provide a quantitative understanding of how these factors affect the stability of the HBs. These differences in $E_a$ between the crystallographic phases (orthorhombic, tetragonal, and cubic) and the types of HBs (N—H⋯Br vs. C—H⋯Br) underline the importance of structural composition and bonding chemistry in determining the thermal stability of hydrogen bonds in MAPbBr$_3$ perovskite.

The differences in the behaviour of HBs between N—H⋯Br and C—H⋯Br can be attributed to two fundamental reasons related to the chemical and structural nature of the HBs formed by nitrogen and carbon, respectively.

- Nitrogen is more electronegative than carbon, meaning that N—H bonds are more polarised than C—H bonds. This greater polarisation of the N—H bond allows for a stronger electrostatic attraction between the hydrogen (partially positive) and the bromine atom (partially negative) in the HB, resulting in N—H⋯Br hydrogen bonds generally being stronger than C—H⋯Br bonds.

- N─H···Br bonds are shorter than C─H···Br bonds since an $NH_3^+$ group has one uncompensated proton and undergoes ionic interaction with $Br^-$. Shorter and stronger HBs are generally more stable and have higher dissociation energies. However, the activation energies of N─H···Br bonds are higher only at low temperatures.

Through the CDFs, it has been possible to verify that the geometric conditions that characterize these HBs can be set constant across the entire temperature range studied, i.e., $d(Br - H) < 3$ Å, and $135º < \sphericalangle(Br - H - N) < 180°$ for N─H···Br, and $d(Br - H) < 4$ Å, and $135º < \sphericalangle(Br - H - C) < 180°$ for C─H···Br. For such conditions, the decrease in the maximum count values of the HBs as the temperature increases is noteworthy. With these definitions, it turns out that the MA cations are mostly linked to Br anions by two or three N─H···Br and C─H···Br bonds. For low temperature, three HBs of each type is the most frequent case, decreasing with increasing temperature. For high temperature, two N─H···Br bonds are the most frequent case, with a crossover near 175 K. In contrast, threefold C─H···Br bonding is the most frequent case for all temperatures.

The power spectra show a redshift with increasing temperature for the signals associated to C─H stretching modes (~3000-3200 cm$^{-1}$) and for other modes with lower wavenumber. The same trend is observed for the N-H stretching modes (~3300 cm$^{-1}$), jointly with a strong broadening, which is a signature of HB. Although C─H···Br bonds have lifetimes and activation energies similar to those of N─H···Br bonds, they have no clear effect on vibrational properties at any temperature. Hence, the mere existence of C─H···Br bonds is not due to chemical interaction, but it is probably due to the confinement of MA+ in the cuboctahedral cage of the perovskite structure.

**Acknowledgements**

The authors would like to thank the computing time provided by CICA (*Centro Informático Científico de Andalucía*, https://www.cica.es/) and by the «*Servicio de Supercomputación de la Universidad de Granada*» (https://supercomputacion.ugr.es). We are grateful to the UK Materials and Molecular Modelling Hub for computational resources, which is partially funded by EPSRC (EP/T022213/1, EP/W032260/1 and EP/P020194/1, this work also used the ARCHER2 UK National Supercomputing Service (http://www.archer2.ac.uk)

# SUPPORTING INFORMATION

# Picosecond lifetimes of hydrogen bonds in the halide perovskite CH$_3$NH$_3$PbBr$_3$

Alejandro Garrote-Márquez,[a] Lucas Lodeiro,[b] Norge Cruz Hernández,[a] Xia Liang,[c] Aron Walsh,[c] and Eduardo Menéndez-Proupin[a*]

[a] *Departamento de Física Aplicada I, Escuela Politécnica Superior, Universidad de Sevilla, Seville E-41011, Spain*

[b] *Departamento de Química, Facultad de Ciencias, Universidad de Chile, Las Palmeras 3425, Ñuñoa 7800003, Santiago, Chile*

[c] *Thomas Young Centre and Department of Materials, Imperial College London, London SW7 2AZ, UK*

*Corresponding author. E-mail: emenendez@us.es

# Combined distribution functions of X-H and H···Y (X=N and C, Y=Br) distances in MAPbBr$_3$

A little of additional information is provided by the CDFs of distances H—Br and X—H. The CDFs for temperatures 100 K and 350 K, typical of the orthorhombic and cubic phases, are shown in Figure S1. At 100 K, there is a characteristic "island" of low H—Br distance, a signature of the N—H···Br bond. That kind of "island" does not appear for the C—H···Br bond. The CDFs are broadened at 350 K, and the "island" disappears.

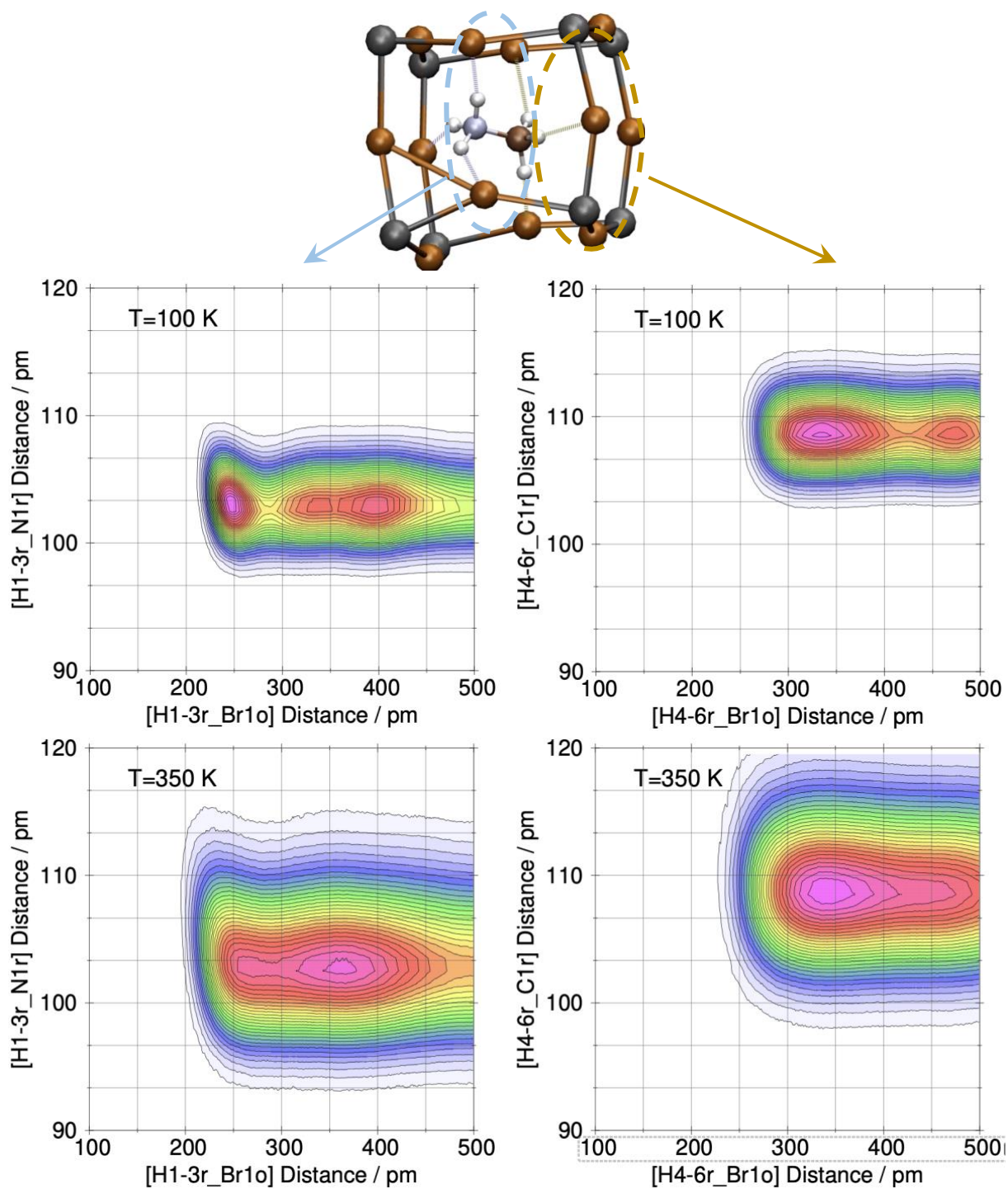

Figure S1. Combined distribution functions of the H-Br distance with either H-N or H-C distance.

Let us see the information given by the correlation functions of the same distances (see J. Chem. Phys. 152, 164105 (2020); doi: 10.1063/5.0005078). These are computed by the TRAVIS code, by subtracting from the CDFs, the Cartesian product the histograms of the two distances. Positive values indicate that the probability of finding this configuration is larger than if the two quantities would be uncorrelated, while negative values depict the opposite situation. Figure S2 shows that configurations with shortened H—Br distance and enlarged X—H distance present positive correlation. This correlation reveals both kinds of HBs at both temperatures.

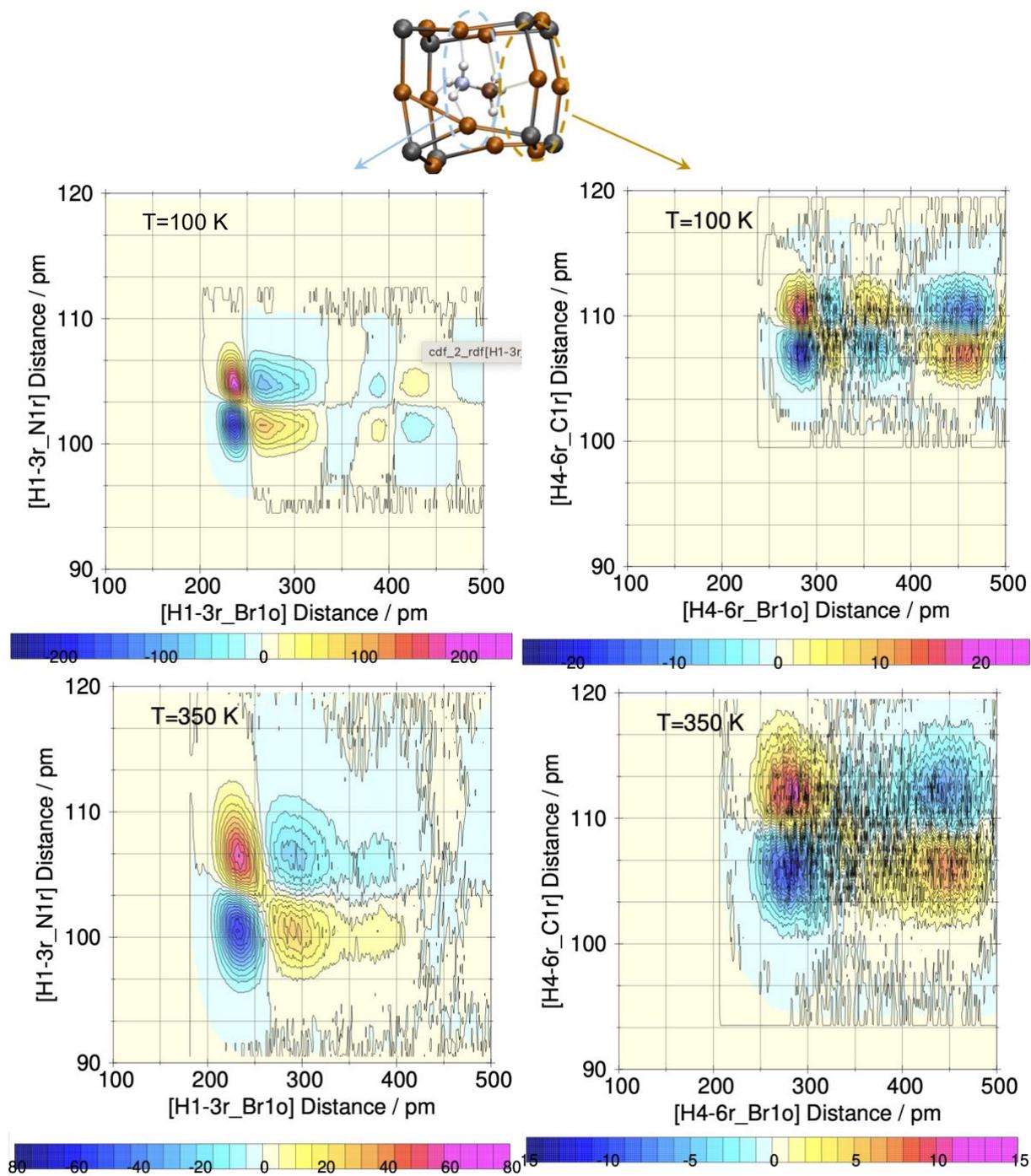

Figure S2. Correlation plot of distances X—H and H—Br in N—H···Br and C—H···Br bonds.

## Power spectrum of MAPbBr$_3$ and the choice of thermostat mass (SMASS)

To ensure that the thermostat does not interfere with the dynamic of the hydrogen bonds, we have explored the effect of the Nose thermostat mass on the dynamic properties, which in the VASP code is regulated by the parameter SMASS. This parameter is automatically set in VASP so that the oscillation of the temperature has a period of 40 MD steps. In our system, that automatic value is SMASS =0.04. With the time step of 0.5 fs, the oscillation period is 20 fs, corresponding to a wavenumber of 1668 cm$^{-1}$. This is very close to the centre of the power spectrum of MAPbBr$_3$ (see Figure S3 below) and should be optimal for efficient thermalization, but it may interfere with the cation dynamics, and it may affect the computed lifetimes. Hence, we tried with SMASS=0.2 and SMASS=1.0, the latter corresponding with an oscillation of temperature every 100 fs or a wavenumber of 334 cm$^{-1}$, which is in the range of normal modes of the inorganic sublattice. As Figure S3 shows, the differences in the power spectrum are negligible.

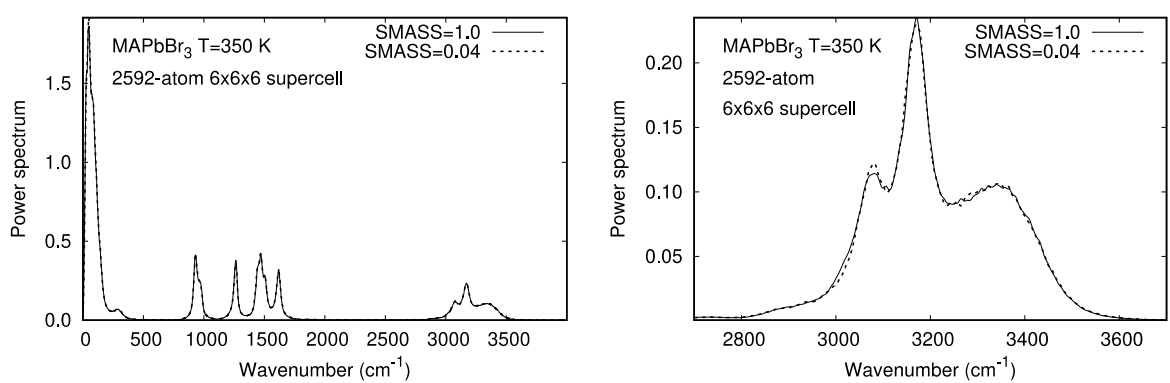

Figure S3. Power spectrum de MAPbBr$_3$. Comparison of the power spectrum for two different thermostat masses (VASP parameter SMASS).

Figure S4 shows that the N─H···Br bond existence autocorrelation function at 350 K. As before, the effect of the SMASS parameter and the simulation time is small.

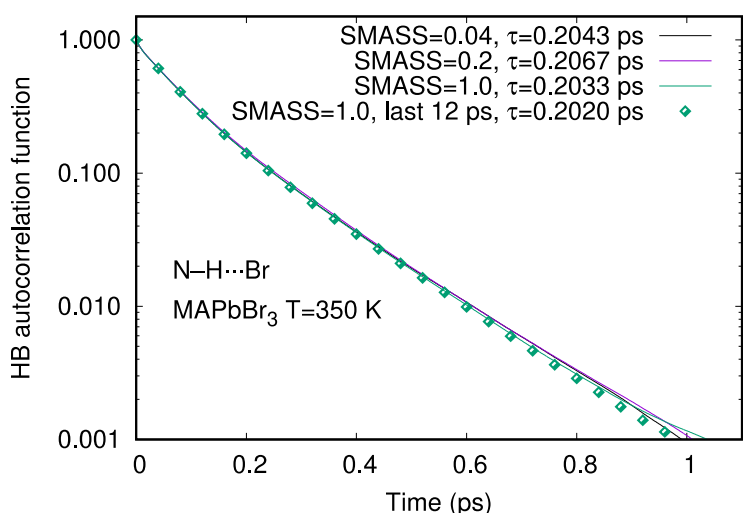

Figure S4. HB existence autocorrelation function of MAPbBr$_3$ at 350 K, computed with different thermostat mass parameters (SMASS), and with different simulation lengths (20 ps if not specified.). The HB lifetimes τ are indicated.

Figure S5 shows several indications of the quality of the MD simulation. In part (a) is the trace of the potential energy, the total energy (potential+kinetic) and the conserved quantity in the

NPT dynamics. Parts (b-f) show the velocity distribution function for each element, and the temperature obtained by fitting with the Maxwell distribution.

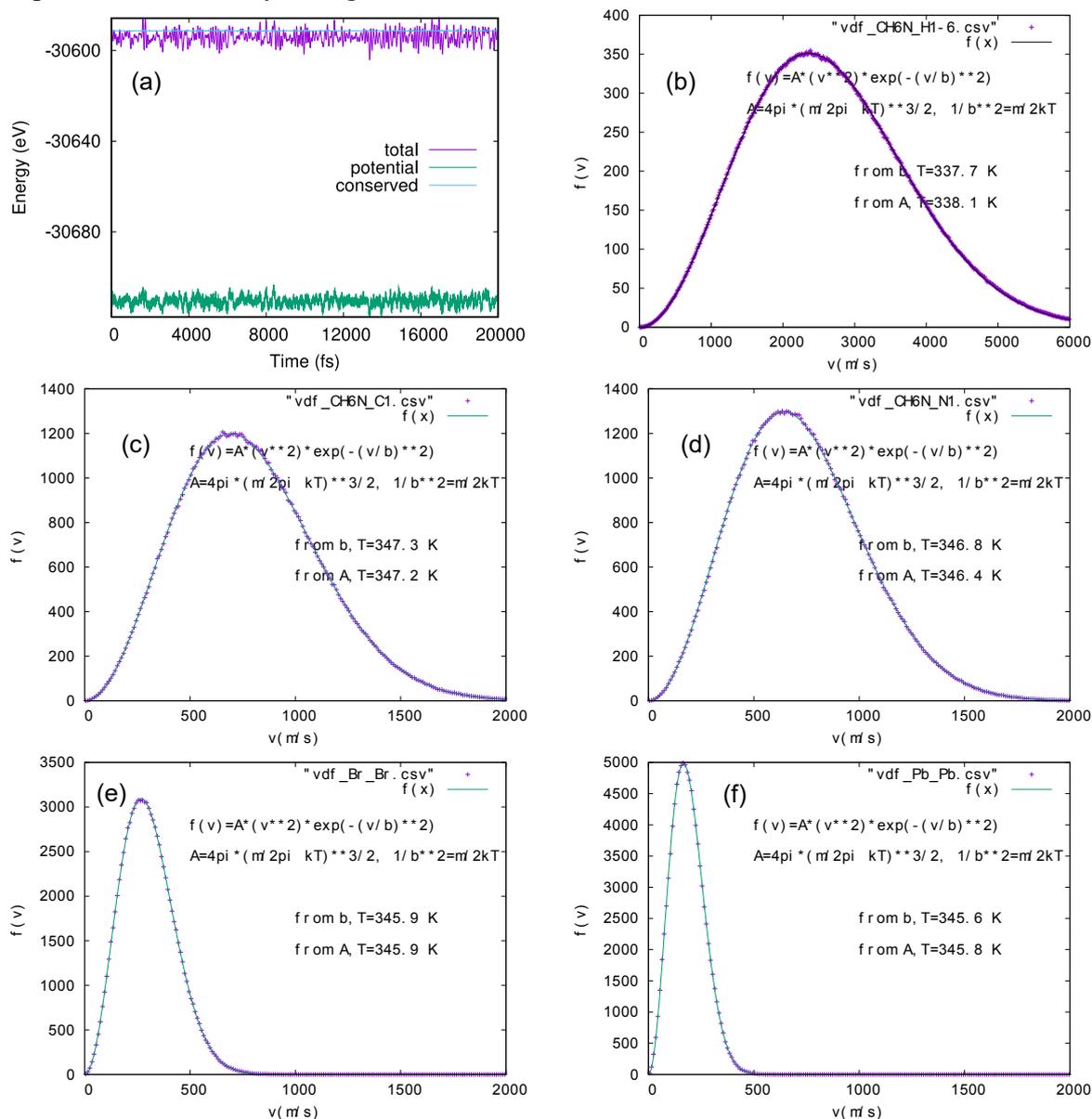

Figure S5. Some results from the MD simulation at 350 K. a) Energies along the simulation time: potential energy, total (potential+kinetic), and the conserved energy, which is the Nose Hamiltonian. b-f) velocity distribution function for each species, fits to the Maxwell distribution functions, and the temperatures derived from the standard deviation and from the pre-exponential factor.

## Neighbor analysis

The following information appears in files cond*.txt and travis.log, after a DACF calculation. For example, for temperature 350 K we obtain
 - 0 Neighbors: 2.5753 percent of the time (111255 hits).
 - 1 Neighbors: 18.2884 percent of the time (790061 hits).
 - 2 Neighbors: 50.7102 percent of the time (2190681 hits).
 - 3 Neighbors: 28.1628 percent of the time (1216635 hits).
 - 4 Neighbors: 0.2620 percent of the time (11318 hits).
 - 5 Neighbors: 0.0012 percent of the time (50 hits).

Based on the above numbers, one can say that on average each MA links with Br through 1 HB 18.3% of the time, 2 HBs 50.7%, 3 HBs 28.2% and makes no HB 2.6% of the time. 4 HBs also appear 0.26% of the time, they can be things like an H linked to 2Br and the other 2 H to two other Br, or 2H with 2Br each, or an H with 3 Br and the other with a Br. 0.26 % is little, but there are 11318 hits, it seems like a good statistic. The same analysis was done for all other simulated temperatures, and it is summarized in Figure S6.

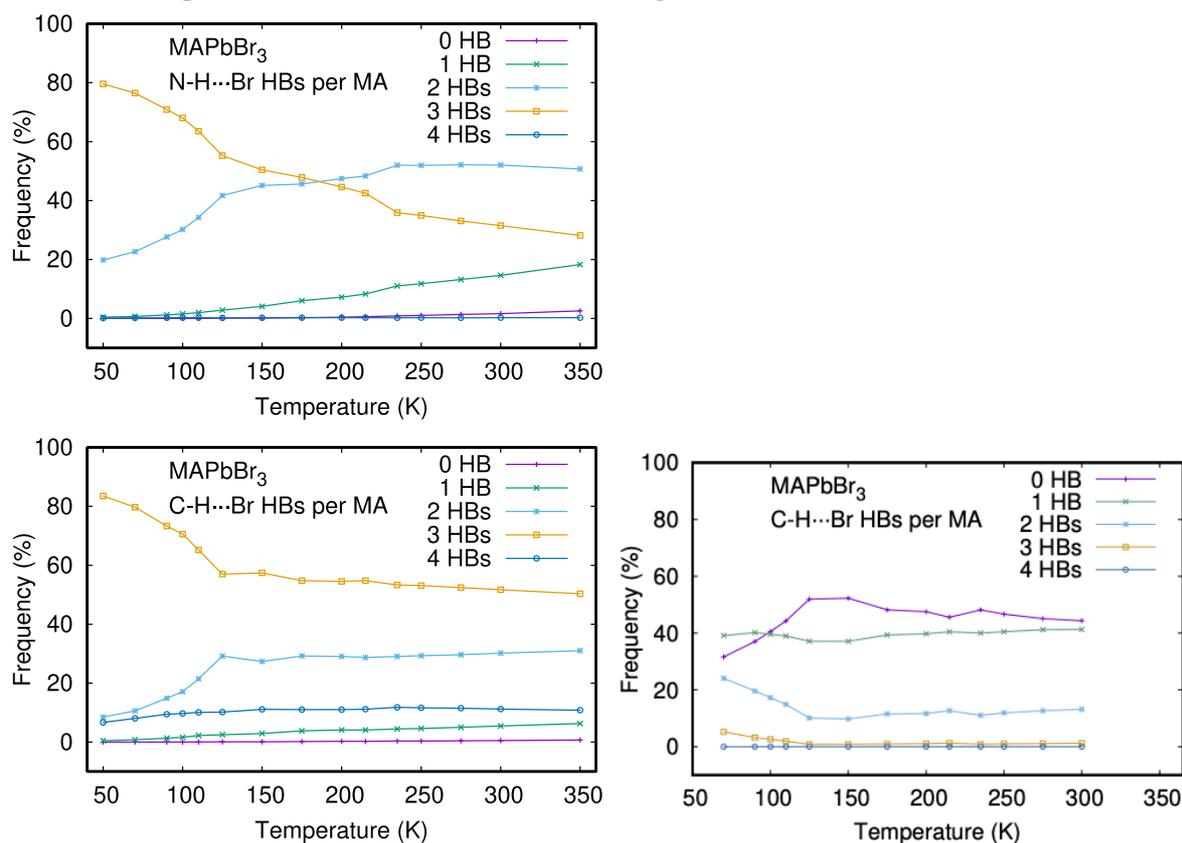

Figure S6. Distribution of the number of bromine ions linked to MA cations through N─H⋯Br or C─H⋯Br HBs, as function of temperature. At the bottom we see to versions of the C-H…Br statistics, with cutoff distance of 400 pm (left) and 300 pm (right).

For 350 K we have evaluated the effect of the SMASS Parameter. The largest variation is 0.2%.

|  | SMASS=1.0 | SMASS=0.04 |
|---|---|---|
| - 0 Neighbors: | 2.5753% | 2.4983% |
| - 1 Neighbors: | 18.2884 | 18.2802% |
| - 2 Neighbors: | 50.7102 | 50.5972% |
| - 3 Neighbors: | 28.1628 | 28.3776% |
| - 4 Neighbors: | 0.2620 | 0.2461% |
| - 5 Neighbors: | 0.0012 | 0.0005% |

**Effect of the correlation depth on the computed lifetimes**

Figure S7 shows one of the HB autocorrelation functions for different correlation depths. The correlation depth, parameter of the TRAVIS code, is the maximum time the autocorrelation function is computed for, and this affects the computed HB lifetimes, which are indicated in the Figure. It is shown that the default parameter, 18 ps for a simulation time of 60 ps, has an excellent convergence.

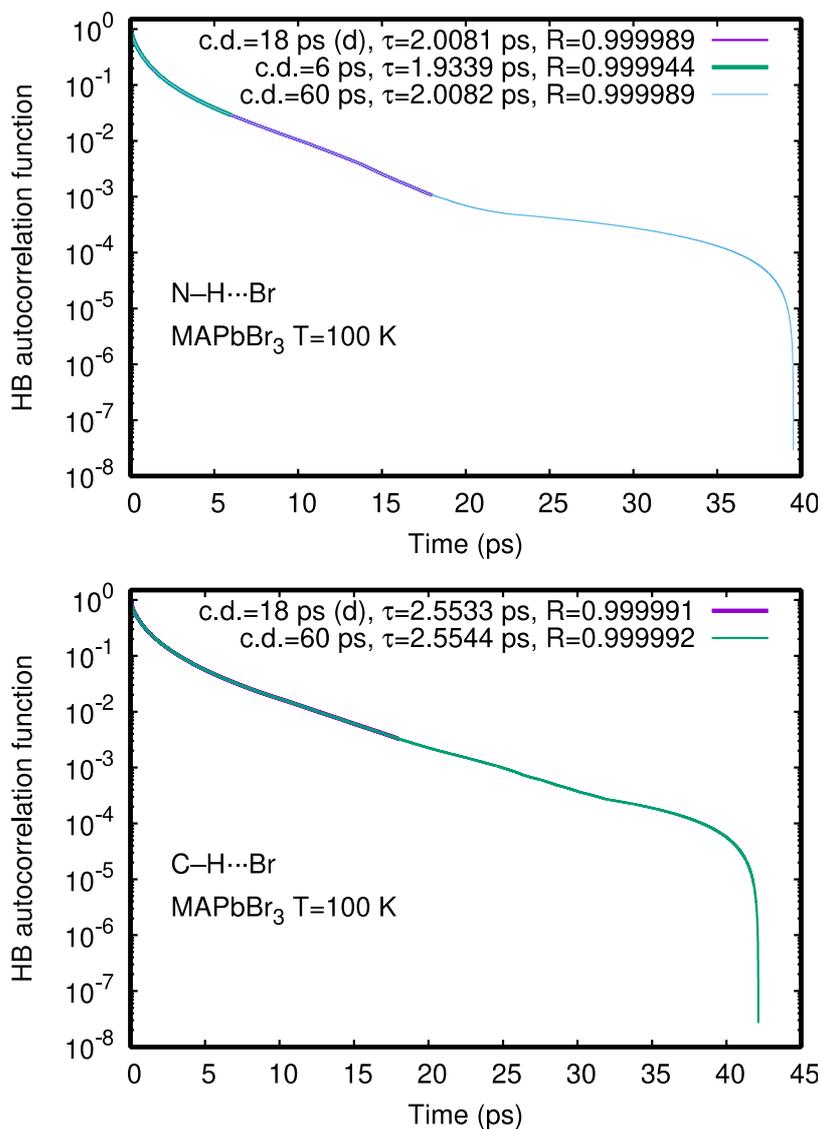

Figure S7. Effect of the correlation depth. The label (d) indicates the value suggested by the TRAVIS code.

**Arrhenius vs Eyring plots**

We have also analysed the LT of the HBs using the Eyring equation, checking which equation best models the behaviour of the LT of the HBs. The Eyring equation is expressed as:

$$k = \frac{k_B T}{h} e^{-\frac{\Delta g}{k_B T}}.$$

In this equation $k = 1/\tau$ represents the rate constant of the reaction. $k_B$ is the Boltzmann constant, $h$ is the Planck constant, $\Delta g$ is the Gibbs free energy of activation per molecule. The Eyring equation can be recast in linearized form as

$$\ln\left(\frac{1}{\tau T}\right) = -\frac{\Delta h}{k_B}\frac{1}{T} + \ln\left(\frac{k_B}{h}\right) + \frac{\Delta s}{k_B},$$

where $\Delta g = \Delta h - T\Delta s$, has been expressed by the enthalpy and entropy $\Delta h$ and $\Delta s$, respectively.

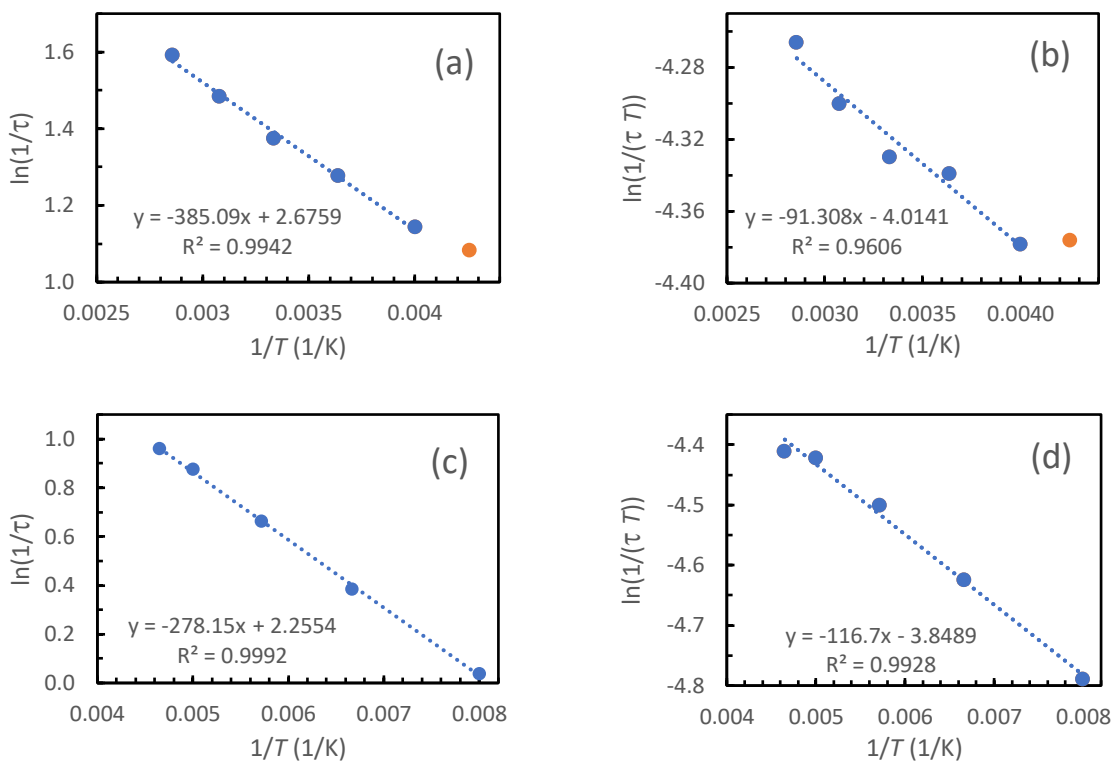

Figure S8. Comparison of linear fits with the Arrhenius a,c) equation and the Eyring equation (b,d) for the N—H···Br bonds. The top row and bottom correspond to the cubic and tetragonal phases (c,d). The point of 235 K (in orange colour) was not included in the fit.

Figure S8 above shows that the fit to the Eyring equation is worse than the fit to the Arrhenius equation.